\documentclass[draftcls,12pt,onecolumn,dvipdfm]{IEEEtran}

\usepackage{mathrsfs}
\usepackage{amsmath}
\usepackage{amssymb}
\usepackage{enumerate}
\usepackage{color,xcolor}
\usepackage{graphicx}
\usepackage{subfigure}
\usepackage{pifont}

\ifCLASSINFOpdf

\else

\fi

\newtheorem{theorem}{Theorem}
\newtheorem{proposition}{Proposition}

\newtheorem{lemma}{Lemma}
\newtheorem{algorithm}{\textbf{Algorithm}}
\usepackage[square,numbers,sort&compress]{natbib}

\usepackage{ifpdf}

\begin{document}

\title{A New Simulation Approach to Performance Evaluation of Binary
Linear Codes in the Extremely Low Error Rate Region}

\author{Xiao~Ma,~\IEEEmembership{Member,~IEEE,}
Jia Liu, and Shancheng Zhao,~\IEEEmembership{Member,~IEEE}
\thanks{This work was presented in part at ISTC'2016.}
\thanks{This work was supported by the NSF of China~(No.~91438101, No.~61401525 and No.~61501206) and the Training Program for Outstanding Young Teachers in Higher Education Institutions of Guangdong Province~(No.~YQ2015092).}
\thanks{X.~Ma is with the School of Data and Computer Science, Sun Yat-sen University, Guangzhou 510275, China. (E-mail: maxiao@mail.sysu.edu.cn).}
\thanks{J.~Liu is with the School of Information Science and Technology, Zhongkai University of Agriculture and Engineering, Guangzhou 510225, China.  (E-mail: ljia2@mail2.sysu.edu.cn).}
\thanks{S.~Zhao is with the College of Information Science and Technology, Jinan University, Guangzhou 510632, China. (E-mail: shanchengzhao@jnu.edu.cn).}
 }


\maketitle

\begin{abstract}
In this paper, the sphere bound~(SB) is revisited within a general bounding framework based on nested Gallager regions. The equivalence is revealed between the SB proposed by Herzberg and Poltyrev
and the SB proposed by Kasami {\em et~al.}, whereas the latter was rarely cited in the literatures. Interestingly and importantly, the derivation of the SB based on nested
Gallager regions suggests us a new simulation
approach to performance evaluation of binary linear codes over additive white Gaussian
noise (AWGN) channels. In order for the performance evaluation, the proposed approach decouples the geometrical structure of the code from the noise statistics. The former specifies the conditional error probabilities, which are independent of signal-to-noise ratios~(SNRs) and can be  simulated and estimated efficiently, while the latter determines the probabilities of those conditions, which involve SNRs and can be calculated numerically. Numerical results show that the proposed simulation approach matches well with the traditional
simulation approach in the high error rate region but is able to evaluate efficiently the performance in the extremely low error rate region.
\end{abstract}



\IEEEpeerreviewmaketitle

\section{Introduction}\label{introduction}

For certain communication systems, including optical fiber
transmission systems~\cite{Marchant90} and magnetic storage systems~\cite{Sripimanwat05}, it is of importance to design coding schemes with extremely low error rate. Typically, it is extremely time-consuming and even infeasible to evaluate such designs by the traditional {\em Monte Carlo} simulations with a limited computational resource due to the difficulty in sampling
a sufficient number of rare error events. One can rely on field-programmable
gate array (FPGA)-based emulation as exhibited by~\cite{Chen11}, where low-density parity-check (LDPC) codes are evaluated empirically down to bit-error rates (BERs) of $10^{-12}$ and below, or make use of importance sampling as proposed in~\cite{Richardson03} based on the knowledge of trapping sets. One can also use
tight bounds to predict their performance without resorting to computer simulations. As reviewed in~\cite{Sason06}, many previously reported upper bounds~\cite{Kasami92,Kasami93,Sphere94,TSB94,Divsalar99,Ma13,Liu14,Liu15} are based on Gallager's first bounding technique~(GFBT)
\begin{equation}\label{GFBT}
  {\rm Pr} \{E\} \leq {\rm Pr} \{E,\underline{y}\in\mathcal{R}\} + {\rm Pr}\{\underline{y}\notin \mathcal{R}\},
\end{equation}
where $E$ denotes the error event, $\underline{y}$ denotes the received signal vector, and $\mathcal{R}$ denotes an arbitrary region around the transmitted signal vector. The second term at the right-hand side~(RHS) of~(\ref{GFBT}) represents the probability of the event that the received vector $\underline{y}$ falls outside the region $\mathcal{R}$, which is considered to be decoded incorrectly even if it may not fall outside the Voronoi region~\cite{Agrell98} of the transmitted codeword. Typically, the first term at the RHS of~(\ref{GFBT}) can be bounded by the union bound. For convenience, we call~(\ref{GFBT}) the {\em $\mathcal{R}$-bound}. As pointed out in~\cite{Sason06},
the choice of the region $\mathcal{R}$ is very significant, and
different choices of this region have resulted in various different
improved upper bounds. For example,
the region $\mathcal{R}$ can be chosen as an $n$-dimensional sphere with center at the
transmitted signal vector and radius $r$, that is the sphere bound~(SB)~\cite{Sphere94}. Intuitively, the more similar the region $\mathcal{R}$ is to the Voronoi region of the transmitted signal vector, the tighter the $\mathcal{R}$-bound is. Therefore, both the shape and the size of the region $\mathcal{R}$ are critical to GFBT. The key difference among existing bounds lies in the shape of the region $\mathcal{R}$. Given the region's shape, one can optimize its size to obtain the tightest $\mathcal{R}$-bound. In most existing bounds, the optimal size of $\mathcal{R}$ is obtained by setting the partial derivative of the bound with respect to a parameter~(specifying the size) to zero.

However, all existing bounds require knowledge of the whole (or truncated) weight enumerating function (WEF) of the code, and may be loose in the high error rate region. In this paper, we will propose a general bounding framework based on nested Gallager's regions and re-visit the SB within the proposed framework. On one hand, the re-derivation reveals the equivalence between the SB proposed by Herzberg and Poltyrev~\cite{Sphere94} and the SB proposed by Kasami {\em et~al.}~\cite{Kasami92,Kasami93}. The former is often cited, as evidenced by the  tutorial book~\cite{Sason06} by Sason and Shamai where the SB proposed by Herzberg and Poltyrev was reviewed as the only form of the SB. In constrast, the SB proposed by Kasami {\em et~al.} was rarely cited in the literatures. On the other hand, the re-derivation stimulates us to develop a new simulation approach to evaluate the performance of binary linear codes over additive white Gaussian noise~(AWGN) channels. The new approach simulates directly the conditional error probability on the spheres with relatively large radii,
where an important fact is invoked that the noise is uniformly distributed over the sphere. Most importantly, the proposed simulation approach can be used to estimate the number of codewords with minimum  Hamming weight, which is critical to  evaluate the conditional error probability on those small spheres. The proposed simulation approach not only matches well with the traditional
simulation approach in the high error rate region but also is able to evaluate efficiently the performance in the extremely low error rate region.

\textbf{Structure:} The rest of this paper is organized as follows. In
Sec. II, the nested Gallager regions with a single parameter is
introduced to exploit GFBT. The SB is then re-derived in the proposed framework.
In Sec. III, within the proposed framework, an alternative
simulation approach is proposed based on several proved geometrical properties. Numerical examples are provided to confirm the analysis in Sec. IV and
we conclude this paper in Sec. V.

\section{A General Bounding Framework Based on Nested Partition}\label{sec2}

\subsection{The System Model}
Let $\mathscr{C}$$[n,k, d_{\min}]$ be a binary linear block code of dimension
$k$, length $n$, and minimum Hamming distance $d_{\min}$.
Suppose that a codeword $\underline{c}=(c_0,
c_1, \cdots, c_{n-1}) \in \mathscr{C}$ is modulated by binary phase
shift keying~(BPSK), resulting in a bipolar signal vector
$\underline s$ with $s_t = 1 - 2c_t$ for $0\leq t \leq n-1$.
The signal vector $\underline s$ is transmitted over an AWGN channel.
Let $\underline{y} = {\underline s} + {\underline z}$ be the
received vector, where $\underline z$ is
a vector of independent Gaussian random variables with zero mean and
variance $\sigma^2$. For AWGN channels, the
maximum-likelihood (ML) decoding is equivalent to finding the nearest
signal vector $\hat{\underline s}$ to $\underline y$. Without loss
of generality, we assume that the bipolar image $\underline s^{(0)}$ of the all-zero codeword $\underline
c^{(0)}$ is transmitted.

The input output weight enumerating function~(IOWEF) of
$\mathscr{C}$ is defined as~\cite[(2.6)]{Sason06}
\begin{eqnarray}
    A(X,Z) \stackrel{\Delta}{=} \sum_{i,j}A_{i,j}X^{i}Z^{j},
\end{eqnarray}
where $X,Z$ are two dummy variables and $A_{i,j}$ denotes the number
of codewords having Hamming weight $j$ when the input information bits having Hamming weight $i$. Then the WEF $A(Z) \stackrel{\Delta}{=} \sum_{j}A_{j}Z^{j}$,
where $A_j = \sum_{i}A_{i, j}$, $0 \leq j \leq n$, is referred to as
the weight spectrum of the given code $\mathscr{C}$.

\subsection{A General Bounding Framework Based on Nested Regions}
In this subsection, we present a general bounding framework based on nested Gallager regions with parameters. To this end, let $\{\mathcal{R}(r), r \in \mathcal{I} \subseteq\mathbb{R}\}$ be a family of Gallager's regions with the same shape and parameterized by $r\in \mathcal{I}$. For example, the nested regions can be chosen as a family of $n$-dimensional spheres of radius $r \geq 0$ centered at the transmitted codeword ${\underline s}^{(0)}$. We make the following assumptions.
\begin{enumerate}
  \item[A1.] The regions $\{\mathcal{R}(r), r \in \mathcal{I} \subseteq\mathbb{R}\}$  are {\em nested} and their boundaries partition the whole space $\mathbb{R}^n$. That is,

       \begin{equation}\label{Assump1-1}
            \mathcal{R}(r_1) \subset \mathcal{R}(r_2)~{\rm if}~r_1  < r_2,
       \end{equation}

       \begin{equation}\label{Assump1-2}
            \partial\mathcal{R}(r_1) \bigcap \partial\mathcal{R}(r_2)=\emptyset~{\rm if}~r_1 \neq r_2,
       \end{equation}

       and

       \begin{equation}\label{Assump1-3}
            \mathbb{R}^n = \bigcup\limits_{r \in \mathcal{I}} \partial\mathcal{R}(r),
        \end{equation}
        where $\partial\mathcal{R}(r)$ denotes the boundary
        surface of the region $\mathcal{R}(r)$.

  \item[A2.] Define a functional $R: {\underline y} \mapsto r$ whenever ${\underline y} \in \partial\mathcal{R}(r)$. The randomness of the received vector $\underline y$ then induces a random variable $R$. We assume that $R$ has a probability density function~(pdf) $g(r)$.

  \item[A3.] We also assume that the conditional error probability ${\rm Pr}\{E|{\underline y} \in \partial\mathcal{R}(r)\}$ can be upper-bounded by a computable upper bound $f_u(r)$.
\end{enumerate}

For ease of notation, we may enlarge the index set $\mathcal{I}$ to $\mathbb{R}$ by setting $g(r) \equiv 0$ for $r\notin \mathcal{I}$.
Under the above assumptions, we have,
\begin{equation}\label{PGFBT2}
  {\rm Pr} \{E\} = \int_{-\infty}^{+\infty} {\rm Pr}\{E|{\underline y} \in \partial\mathcal{R}(r)\} g(r)~{\rm d}r.
\end{equation}

Given a binary linear code $\mathscr{C}$, we have the following parameterized GFBT.
\begin{proposition}\label{Proposition_PGFBT}
Let $f_u(r)$ be an upper bound on the conditional error probability ${\rm Pr}\{E|{\underline y} \in \partial\mathcal{R}(r)\}$. For any $r^*\in \mathbb{R}$,
\begin{equation}\label{PGFBT}
  {\rm Pr} \{E\} \leq \int_{-\infty}^{r^*} f_u(r) g(r)~{\rm d}r + \int_{r^*}^{+\infty}g(r)~{\rm d}r.
\end{equation}
\end{proposition}
\begin{IEEEproof}
\begin{eqnarray}
  {\rm Pr} \{E\} &=&  {\rm Pr}\{E, \underline y \in \mathcal{R}(r^*)\} + {\rm Pr}\{E, \underline y \notin \mathcal{R}(r^*)\}\nonumber\\
              &\leq&  {\rm Pr}\{E, \underline y \in \mathcal{R}(r^*)\} + {\rm Pr}\{\underline y \notin \mathcal{R}(r^*)\}\nonumber\\
                 &=&  \int_{-\infty}^{r^*} f_u(r) g(r)~{\rm d}r + \int_{r^*}^{+\infty}g(r)~{\rm d}r. \nonumber
\end{eqnarray}
\end{IEEEproof}

Given $f_u(r)$, an immediate question is how to choose $r^*$ to make the above bound as tight
as possible. This can be solved by setting the derivative of~(\ref{PGFBT}) with respect to $r^*$ to be zero. This can also be solved by finding the solution of $f_u(r) = 1$, as justified by the following proposition.

\begin{proposition}\label{Proposition_r1}
Assume that $f_u(r)$ is a non-decreasing and continuous function of $r$. Let $r_1$ be a parameter that minimizes the upper bound as shown in~(\ref{PGFBT}). Then, if $f_u(r) < 1$ for all $r \in  \mathcal{I}$, $r_1 = \sup\{r\in\mathcal{I}\}$; otherwise, if $f_u(r) \geq 1$ for some $r  \in  \mathcal{I}$, $r_1$ can be taken as any solution of $f_u(r) = 1$. Furthermore, if $f_u(r)$ is strictly increasing and continuous in an interval $[r_{\min}, r_{\max}]$ such that $f_u(r_{\min}) < 1$ and $f_u(r_{\max}) > 1$, there exists a unique $r_1 \in [r_{\min}, r_{\max}]$ such that $f_u(r_1) = 1$.
\end{proposition}

\begin{IEEEproof}
The second part is obvious since the function $f_u(r)$ is strictly increasing and continuous, while the first part can be proved by arguing that any $r^*$ with $f_u(r^*) \neq 1$ cannot be the minimizer.

Let $r_0 < \sup\{r\in\mathcal{I}\}$ such that $f_u(r_0) < 1$. Since $f_u(r)$ is continuous and $r_0 < \sup\{r\in\mathcal{I}\}$, there exists $\mathcal{I}\ni r' > r_0$ such that $f_u(r') < 1$. Then
\begin{eqnarray}
\lefteqn{\int_{-\infty}^{r_0} f_u(r) g(r)~{\rm d}r + \int_{r_0}^{+\infty}g(r)~{\rm d}r} \nonumber\\
&\!\!\!\!\!\!\!\!\!\!\! \!\!\!=&\!\!\!\!\! \!\!\!\int_{-\infty}^{r_0} f_{u}(r)g(r)~{\rm d}r +\!\!\!
\int_{r_0}^{r'} g(r)~{\rm d}r + \int_{r'}^{+\infty}g(r)~{\rm d}r \nonumber \\
&\!\!\!\!\!\!\!\!\!\!\! \!\!\!>&\!\!\!\!\! \!\!\!
\int_{-\infty}^{r_0} \!\!f_{u}(r) g(r)~{\rm d}r + \int_{r_0}^{r'}\!f_{u}(r)g(r)~{\rm d}r + \int_{r'}^{+\infty}\!\!\!\!\!g(r)~{\rm d}r \nonumber \\
&\!\!\!\!\!\!\!\!\!\!\! \!\!\!=&\!\!\!\!\!
\!\!\!\int_{-\infty}^{r'}f_{u}(r)g(r) ~{\rm d}r +
\int_{r'}^{+\infty}g(r)~{\rm d}r, \nonumber
\end{eqnarray}
where we have used the fact that $f_u(r) < 1$ for $r\in [r_0, r']$. This shows that $r'$ is better than $r_0$.

Let $r_2$ be a parameter such that $f_u(r_2) > 1$. Since $f_u(r)$ is continuous\footnote{We assume that $f_u(r)$ is a nontrivial upper bound in the sense that there exists some $r$ such that $f_u(r) < 1$.}, there exists $r_1 < r_2$ such that $f_u(r_1) = 1$. Then
\begin{eqnarray}
\lefteqn{\int_{-\infty}^{r_2} f_u(r) g(r)~{\rm d}r + \int_{r_2}^{+\infty}g(r)~{\rm d}r} \nonumber\\
&\!\!\!\!\!\!\!\!\!\!\! \!\!\!=&\!\!\!\!\! \!\!\!\int_{-\infty}^{r_{1}} f_{u}(r)g(r)~{\rm d}r +\!\!\!
\int_{r_{1}}^{r_2} f_{u}(r)g(r)~{\rm d}r + \int_{r_2}^{+\infty}g(r)~{\rm d}r \nonumber \\
&\!\!\!\!\!\!\!\!\!\!\! \!\!\!>&\!\!\!\!\! \!\!\!
\int_{-\infty}^{r_1} f_{u}(r) g(r)~{\rm d}r + \int_{r_1}^{r_2}g(r)~{\rm d}r + \int_{r_2}^{+\infty}\!\!\!\!\!g(r)~{\rm d}r \nonumber \\
&\!\!\!\!\!\!\!\!\!\!\! \!\!\!=&\!\!\!\!\!
\!\!\!\int_{-\infty}^{r_1}f_{u}(r)g(r) ~{\rm d}r +
\int_{r_1}^{+\infty}g(r)~{\rm d}r, \nonumber
\end{eqnarray}
where we have used a condition that $f_u(r) > 1$ for $r\in (r_1, r_2]$, which can be fulfilled by choosing $r_1$ to be the maximum solution of $f_u(r) = 1$. This shows that $r_1$ is better than $r_2$.
\end{IEEEproof}

{\bf Remark.~}In contrast to the proof by setting the derivative to be zero, the above proof is more insightful, which actually suggests a more compact form of the GFBT based on nested regions $\mathcal{R}(r)$, as given by
\begin{eqnarray}
    {\rm Pr}\{E\}& \leq & \min_{r^*} \left\{ \int_{-\infty}^{r^*}\!\!\! f_u(r) g(r)~{\rm d}r + \int_{r^*}^{+\infty}\!\!\! g(r)~{\rm d}r \right\}\label{optimal r1}\\
     &=& \int_{-\infty}^{+\infty} \min\{f_u(r), 1\}g(r)~{\rm d}r.
\end{eqnarray}



\subsection{Conditional Union Bound}
We focus on the conditional union bound
\begin{eqnarray} \label{CUnionbound}
    {\rm Pr}\{E|{\underline y} \in \partial\mathcal{R}(r)\} & \leq &f_u(r) = \sum_{1\leq d \leq n} A_d p_2(r, d) ,
\end{eqnarray}
where $\{A_d, 1\leq d \leq n\}$ is the weight spectrum of the code $\mathscr{C}$, and $p_2(r, d)$ is the conditional pair-wise error probability conditional on the event $\{{\underline y} \in \partial \mathcal{R}(r)\}$. We have by definition that
\begin{eqnarray}\label{Cpairwise}
    p_2(r, d) &=& {\rm Pr}\left\{\|\underline{y}-\underline s^{(1)}\|\leq \|\underline{y}-\underline
  s^{(0)}\| \mid \underline{y}\in \partial\mathcal{R}(r)\right\}\nonumber\\
  &=& \frac{\int_{\|\underline{y}-\underline s^{(1)}\| \leq \|\underline{y}-\underline
  s^{(0)}\|,~~\underline{y}\in \partial\mathcal{R}(r)}p(\underline y)~{\rm d}{\underline y}}{\int_{\underline{y}\in \partial\mathcal{R}(r)}p(\underline y)~{\rm d}{\underline y}},
\end{eqnarray}
where $p({\underline y})$ is the pdf of $\underline y$. Noticing that, different from the unconditional pair-wise error probabilities, $p_2(r, d)$ may be zero for some $r$.


We have the following lemma.
\begin{lemma}\label{LemmaCpairwise}
Suppose that, conditional on $\underline{y}\in \partial\mathcal{R}(r)$, the received vector $\underline y$ is uniformly distributed over
$\partial\mathcal{R}(r)$. Then the conditional pair-wise error probability $p_2(r, d)$ does not depend on the signal-to-noise ratio~(SNR).
\end{lemma}
\begin{IEEEproof} Since $p(\underline y)$ is constant for ${\underline y} \in \partial\mathcal{R}(r)$, we have, by canceling $p(\underline y)$ from both the numerator and the denominator of $(\ref{Cpairwise})$,
\begin{equation}\label{Cpairwise1}
    p_2(r, d) = \frac{\int_{\|\underline{y}-\underline s^{(1)}\| \leq \|\underline{y}-\underline
  s^{(0)}\|, \underline{y}\in \partial\mathcal{R}(r)}~{\rm d}{\underline y}}{\int_{\underline{y}\in \partial\mathcal{R}(r)}~{\rm d}{\underline y}},
\end{equation}
which shows that the conditional pair-wise error probability can be represented as a ratio of two ``surface area" and hence does not depend on the SNR.
\end{IEEEproof}

Given that the received vector $\underline y$ is uniformly distributed over $\partial\mathcal{R}(r)$,  $f_u(r)$ in (\ref{CUnionbound}) does not depend on the SNR. Furthermore, the minimizer $r_1$ of (\ref{optimal r1}), as a solution of
$ \sum_{1\leq d \leq n} A_d p_2(r, d) = 1$ (see Proposition~\ref{Proposition_r1}) , does not depend on the SNR, either.

\subsection{Sphere Bound Revisited}\label{sec3}

In this subsection, sphere bound is revisited based on the general framework. Without loss of generality, we assume that the code $\mathscr{C}$ has at least three non-zero codewords, i.e., its dimension $k>1$.
Let ${\underline c}^{(1)}$~(with bipolar image ${\underline s}^{(1)}$) be a codeword of Hamming weight $d$. The Euclidean distance between $\underline s^{(0)}$ and $\underline s^{(1)}$ is $2\sqrt{d}$.

\subsubsection{Nested Regions}

The nested region in sphere bound is the family of $n$-dimensional spheres centered at the transmitted signal vector, that is, $\mathcal{R}(r) = \{\underline{y}\mid \|\underline{y}-\underline s^{(0)}\| \leq r\}$, where $r\geq 0$ is the parameter.

\subsubsection{Probability Density Function of the Parameter}

The pdf of the parameter is
\begin{equation}\label{SB_g}
g(r) = \frac{2r^{n-1}e^{-\frac{r^{2}}{2\sigma^{2}}}}{2^{\frac{n}{2}}\sigma^{n}\Gamma(\frac{n}{2})},~~r\geq 0.
\end{equation}

\begin{figure}
\centering
\includegraphics[width=9cm]{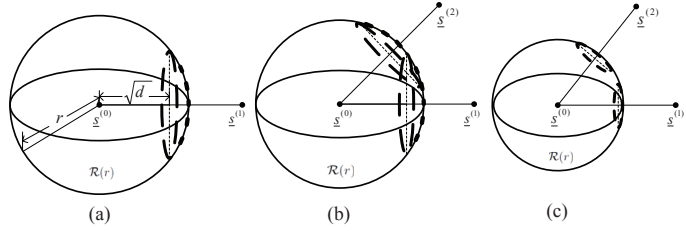}
\caption{Graphical illustration of the conditional error probability on ${\partial\mathcal{R}(r)}.$ } \label{SB_P_T_fig}
\end{figure}

\subsubsection{Conditional Upper Bound}
The SB chooses $f_u(r)$ to be the conditional union bound. Given that $||{\underline y} - {\underline s}^{(0)}|| = r$, $\underline y$ is uniformly distributed over $\partial\mathcal{R}(r)$. Hence the conditional pair-wise error probability $p_2(r, d)$ does not depend on the SNR and can be evaluated as the ratio of the surface area of a spherical cap to that of the whole sphere, as illustrated in Fig. \ref{SB_P_T_fig}~(a). That is,
\begin{equation}\label{SB_f2}
    \!\!\!\!\!p_2(r, d)\!\! = \!\!\left\{\begin{array}{rl}
                                         \frac{\Gamma(\frac{n}{2})}{\sqrt{\pi}~\Gamma(\frac{n-1}{2})}\int_{0}^{\arccos(\frac{\sqrt{d}}{r})}\sin^{n-2}
\phi~{\rm d}\phi, & r > \sqrt{d} \\
                                           0, & r \leq \sqrt{d}
                                         \end{array}\right.,
\end{equation}
which is a non-decreasing and continuous function of $r$ such that $p_2(0, d) = 0$ and $p_2(+\infty, d) = 1/2$. Therefore, the conditional union bound of (\ref{CUnionbound}) is also an non-decreasing and continuous function of $r$ such that $f_u(0) = 0$ and $f_u(+\infty) \geq 3/2$. Furthermore, $f_u(r)$ is a strictly increasing function in the interval $[\sqrt{d_{\min}}, +\infty)$ with $f_u(\sqrt{d_{\min}}) = 0$. Hence there exists a unique $r_1$ satisfying

\begin{equation}\label{SB_opt}
    \sum_{1\leq d \leq n} A_d p_2(r, d) = 1,
\end{equation}
which is equivalent to that given in~\cite[(3.48)]{Sason06} by noticing that $p_2(r, d) = 0$ for $d > r^2$.

\subsubsection{Equivalence}

The SB can be written as
\begin{eqnarray}\label{SB}
  {\rm Pr} \{E\}&\leq& \int_{0}^{r_{1}}f_{u}(r) g(r)  ~{\rm d}r +
\int_{r_{1}}^{+\infty}g(r)~{\rm d}r\nonumber\\
&=& \int_{0}^{+\infty} \min\{f_{u}(r), 1\} g(r) ~{\rm d}r,
\end{eqnarray}
where $f_u(r)$ and $g(r)$ are given in~(\ref{CUnionbound}) and~(\ref{SB_g}), respectively. The optimal parameter $r_{1}$ is
given by solving the equation~(\ref{SB_opt}). It can be seen that~(\ref{SB}) is exactly the sphere
bound of Kasami {\em et al}~\cite{Kasami92}\cite{Kasami93}. It can
also be checked from two aspects that~(\ref{SB}) is equivalent to that given
in~\cite[(3.45)-(3.48)]{Sason06}. First, we have shown that the optimal radius $r_1$
satisfies~(\ref{SB_opt}), which is equivalent to that given
in~\cite[(3.48)]{Sason06}. Second, by changing variables, $z_{1} =
r\cos\phi$ and $y = r^{2}$, it can be verified that~(\ref{SB}) is
equivalent to that given in~\cite[Sec.3.2.5]{Sason06}.

%

\section{A New Simulation Approach}\label{A New Simulation Approach}

The sphere bound, like other bounds, is typically useful for estimating the performance in the high SNR region, where the simulation becomes pale due to the extremely low error rates. However, the sphere bound requires the knowledge of WEF, which is usually not available for a general code. Interestingly and importantly, the derivation of the SB based on nested regions stimulates us to develop a new simulation approach.

In the following, we assume that $\mathcal{R}(r)$ is an $n$-dimensional ball centered at the transmitted signal vector with radius $r$.
We have the following lemma.

\begin{lemma}\label{LemmaCEP}
 The conditional error probability ${\rm Pr}\{E|{\underline y} \in \partial\mathcal{R}(r)\}$ under ML decoding, which vanishes for $r < \sqrt{d_{\min}}$,  does not
depend on the SNR and is an increasing function over $ [\sqrt{d_{\min}}, +\infty)$.
\end{lemma}

\begin{IEEEproof}
Given the code $\mathscr{C}$, the Voronoi region $\mathcal{V}({\underline s}^{(0)})$ of the transmitted signal vector ${\underline s}^{(0)}$ is fixed and contains $\mathcal{R}(r)$ for $r < \sqrt{d_{\min}}$ as subsets, implying that ${\rm Pr}\{E | \underline y\} = 0$ for $r < \sqrt{d_{\min}}$. For $ r \in [\sqrt{d_{\min}}, +\infty)$, the conditional error probability ${\rm Pr}\{E|{\underline y} \in \partial\mathcal{R}(r)\}$ is given by
\begin{eqnarray}\label{CpairwiseSNR}
 {\rm Pr}\{E|{\underline y} \in \partial\mathcal{R}(r)\} = \frac{\int_{\underline{y} \in  \partial\mathcal{R}(r), ~\underline{y} \notin \mathcal{V}({\underline s}^{(0)})}p(\underline y)~{\rm d}{\underline y}}{\int_{\underline{y}\in \partial\mathcal{R}(r)}p(\underline y)~{\rm d}{\underline y}},
\end{eqnarray}
where $p({\underline y})$ is the pdf of $\underline y$. Since the received vector $\underline y$ is uniformly distributed over
$\partial\mathcal{R}(r)$, we have, by canceling $p(\underline y)$ from both the numerator and the denominator of $(\ref{CpairwiseSNR})$,
\begin{eqnarray}
{\rm Pr}\{E|{\underline y} \in \partial\mathcal{R}(r)\} = \frac{\int_{\underline{y} \in  \partial\mathcal{R}(r), ~\underline{y} \notin \mathcal{V}({\underline s}^{(0)})}~{\rm d}{\underline y}}{\int_{\underline{y}\in \partial\mathcal{R}(r)}~{\rm d}{\underline y}}.
\end{eqnarray}
That is, the conditional error probability is equal to the ratio between the ``surface area" of $\partial\mathcal{R}(r) \backslash \mathcal{V}({\underline s}^{(0)})$ and the ``surface area" of $\partial\mathcal{R}(r)$, implying that the conditional error probability does not depend on the SNR.

It is intuitively correct that the conditional error probability is an increasing function, however, a rigorous proof needs patitioning approximately the sphere surface into a collection of spherical cap, each of which is infinitesimal and has an increasing area ratio of the form (\ref{SB_f2}) as $r$ increases.

\end{IEEEproof}

We have the following theorem, which provides an estimate of the error rate.

\begin{theorem}\label{Theorem_NEW}
Let $f(r) = {\rm Pr}\{E|{\underline y} \in \partial\mathcal{R}(r)\}$,  where $\partial\mathcal{R}(r)$ is  the sphere of radius $r$. For any  $L$ real numbers, $\sqrt{d_{\min}} = r_0<r_1<\cdots < r_{{\ell} - 1} <r_{L}=+\infty$, we have
\begin{equation}\label{NEWBound}
  {\rm Pr} \{E\} \leq \sum_{\ell=1}^{L} p_{\ell} f(r_{\ell}),
\end{equation}
where
\begin{equation}\label{pi}
p_{\ell}= \int_{r_{\ell-1}}^{r_{\ell}}g(r)~{\rm d}r
\end{equation}
and
\begin{equation*}
g(r) = \frac{2r^{n-1}e^{-\frac{r^{2}}{2\sigma^{2}}}}{2^{\frac{n}{2}}\sigma^{n}\Gamma(\frac{n}{2})}
\end{equation*}
 is the pdf of the radius $r$.
%
%
\end{theorem}
\begin{IEEEproof}
From (\ref{PGFBT2}), we have
\begin{eqnarray*}
 {\rm Pr} \{E\} &=& \int_{0}^{+\infty} f(r) g(r)~{\rm d}r\\
   &=&  \sum_{\ell=1}^{L}\int_{r_{\ell-1}}^{r_{\ell}} f(r) g(r)~{\rm d}r\\
   &\leq& \sum_{\ell=1}^{L}\int_{r_{\ell-1}}^{r_{\ell}} f(r_{\ell}) g(r)~{\rm d}r\\
   &=& \sum_{\ell=1}^{L} p_{\ell} f(r_{\ell}),
\end{eqnarray*}
where the inequality follows from Lemma~\ref{LemmaCEP}.

\end{IEEEproof}

From Theorem~\ref{Theorem_NEW}, the frame error rate (FER) can be approximated as the partition of the interval $ [\sqrt{d_{\min}}, +\infty)$ becomes finer given that the conditional error probabilities $ f(r_{\ell})$ are available for all $\ell$ ($0 \leq \ell \leq L$). One main contribution of this paper is to provide a simulation method to evaluate the conditional error probability.

\begin{algorithm}{\bf \em A Simulation Procedure to Estimate the Conditional Error Probability}\label{alg:new simulation approach}
\begin{enumerate}
  \item Generate an information vector $\underline{u}$ randomly, which is then encoded into a codeword $\underline{c}$ by the encoder of $\mathscr{C}$ and modulated into a  real vector $\underline{s}$;
  \item Generate a noise vector $\underline{w}$ randomly according to the standard Gaussian distribution $\mathcal{N}(0,1)$, which is then normalized as a unit vector $\underline{z}=\frac{\underline{w}}{||\underline{w}||}$;
  \item Add $r \underline{z}$ to $\underline{s}$, resulting in $\underline{y}$, which is uniformly distributed over the sphere $\partial\mathcal{R}(r)$ centered at $\underline{s}$. Given $\underline{y}$, execute a decoding algorithm to obtain an estimate $\hat{u}$ of the transmitted vector.
\end{enumerate}
\end{algorithm}


For large $r$, $f(r) = {\rm Pr}\{E|{\underline y} \in \partial\mathcal{R}(r)\}$ is large and can be estimated reliably by {\em Monte Carlo} simulation with a reasonable computational resource. For example, in the case when $f(r)$ is around $10^{-6}$, we can implement Algorithm~\ref{alg:new simulation approach} around
(say) $10^{9}$ times, obtaining a relative frequency of the decoding error event, which is widely accepted as an accurate~({\em in probability}) estimate of $f(r)$. We can assume that $f(+\infty) = 1$. As $r$ becomes small, the conditional error probability $f(r)$ can be extremely small, making it time-consuming and even infeasible to implement the {\em Monte Carlo} simulation based on Algorithm~\ref{alg:new simulation approach}. One way to resolve this difficulty is to use the conditional union bound, which requires the (truncated) WEF,  as has been done in~\cite{Zhao2016}. Another way, which requires little knowledge of the WEF, is based on the following lemma.

\begin{lemma}\label{LemmaLittleSphere}
There exists a radius $\delta_1 > \sqrt{d_{\min}}$ such that
\begin{equation}\label{f_u(r)_littleframe}
     f(r) = {\rm Pr}\{E|{\underline y} \in \partial\mathcal{R}(r)\} = A_{d_{\min}}{p_2(r, d_{\min}) }
\end{equation}
for $r \in [\sqrt{d_{\min}},  \delta_1]$.
\end{lemma}

\begin{IEEEproof}
This can be proved by noticing the following two facts as $r$ becomes small, see Fig.~\ref{SB_P_T_fig} for an illustration.
\begin{itemize}
\item As $r$  becomes small enough, only codewords with minimum distance have contributions to the conditional error rates.
\item As $r$ becomes small enough, the spherical caps corresponding to different codewords become non-overlapped.
\end{itemize}
\end{IEEEproof}

An important consequence of Lemma~\ref{LemmaLittleSphere} is that
\begin{equation}
     A_{d_{\min}} =  \frac{f(\delta_1)}{p_2(\delta_1, d_{\min}) }.
\end{equation}
Although $\delta_1$ is in general not available, we can still have\footnote{Such an approximation is reasonable since the conditional error probability on small spheres is dominated by the minimum distance.}
\begin{equation}
A_{d_{\min}} \approx \frac{f(r)}{p_2(r, d_{\min}) }
\end{equation}
by estimating $f(r)$ via {\em Monte Carlo} simulation for a small $r$. It is not difficult to imagine that the estimated $A_{d_{\min}}$ varies from simulations even with the same radius $r$ due to the nature of the {\em Monte Carlo} method. One way to mitigate the possible effects caused by the simulation is to have several estimates of $A_{d_{\min}}$ for different small radii and then to take the average into use.
However, we observe that the estimated errors (within a reasonable range) in $A_{d_{\min}}$ have little effects on the performance curve. This can be explained numerically as follows. Suppose that the estimated $\hat{A}_{d_{\min}} = 1050$, in contrast to the true value $A_{d_{\min}} = 1000$. Then the error $|\hat{A}_{d_{\min}} - A_{d_{\min}}| p_2(r, d_{\min})$ can be neglected compared to $A_{d_{\min}} p_2(r, d_{\min})$ since their exponents are not in the same level.

We shall conclude this section by summarizing the new simulation procedure. For comparison, we list the traditional simulation procedure below.

%
%
%

\begin{algorithm}{\bf \em Traditional Simulation Approach}\label{alg:traditional simulation approach}
\begin{enumerate}
  \item Generate an information vector $\underline{u}$ randomly, which is then encoded into a codeword $\underline{c}$ by the encoder of $\mathscr{C}$ and modulated into a  real vector $\underline{s}$;
  \item Generate a noise vector $\underline{w}$ randomly according to the standard Gaussian distribution $\mathcal{N}(0,1)$;
  \item Add $\sigma \underline{w}$ to $\underline{s}$, resulting in $\underline{y}$. Given $\underline{y}$, execute a decoding algorithm to obtain an estimate $\hat{u}$ of the transmitted vector.
\end{enumerate}
\end{algorithm}

One issue associated with the above simulation approach is: to obtain the performance at
another SNR, new simulations have to be executed and all simulation results obtained previously
become useless. Another issue is that, if we are interested in the performance of the code over other memoryless channels with  real additive noise, we need totally different simulations. Let $\delta_2$ be a large real number and $\Delta > 0$ be a small real number. We have the following simulation approach.

\begin{algorithm}{\bf \em New Simulation Approach}\label{alg:new simulation approachforallr}

\begin{enumerate}
  \item For $r=\delta_2, \delta_2-\Delta,\delta_2-2\Delta,\cdots$, estimate $ f(r) = {\rm Pr}\{E|{\underline y} \in \partial\mathcal{R}(r)\}$ by Algorithm~\ref{alg:new simulation approach} up to $r = \delta_1$ such that $f(\delta_1)$ becomes too small to be estimated reliably by simulation in a reasonable time. For example, this step can be terminated at $r = \delta_1$ such that $f(\delta_1) \approx 10^{-6}$;
  \item Estimate $ A_{d_{\min}} \approx  \frac{f(\delta_1)}{p_2(\delta_1, d_{\min}) }$ by Lemma~\ref{LemmaLittleSphere};
  \item For $r= \delta_1-\Delta,\delta_1-2\Delta,\cdots,\delta_1 - m\Delta$, where $ m = \lfloor  \frac{\delta_1 - \sqrt{d_{\min}} }{\Delta} \rfloor$, estimate $ f(r) \approx  A_{d_{\min}} p_2(r, d_{\min})$.
\end{enumerate}
\end{algorithm}

{\bf Remarks:~}The distinguished feature of the new simulation approach is its independence of the SNR. Actually, each simulated $f(r_\ell)$ relies only on the geometrical structure of the coded signals in the Euclidean space. Suppose that the conditional error probabilities $f(r_\ell)$, for $r_{1} = \delta_1 - m\Delta, r_{2} =\delta_1 - (m-1)\Delta,\cdots,r_{L-1} = \delta_2,r_{L} =+\infty$, have been estimated by Algorithm~\ref{alg:new simulation approachforallr}. Then the error probability is obtained by
Theorem~\ref{Theorem_NEW} as
\begin{equation}
{\rm Pr} \{E\} \approx \sum_{\ell=1}^{L} p_{\ell} f(r_{\ell}),
\end{equation}
where the coefficients $p_\ell$ have analytical forms as shown in (\ref{pi}) and are computable for any given SNR. At a first glance, to obtain the conditional error probability on a relatively small sphere, the new simulation approach requires the exact value of the minimum Hamming distance of the considered code. However, we will show later by numerical example that a lower bound on the minimum distance can also be used to obtain an approximated curve. In this case, Lemma~\ref{LemmaLittleSphere} is adapted to
\begin{equation}
 f(r) = A_{d_{\min}} p_2(r, d_{\min}) \leq A_{d_{\min}} p_2(r, \hat d),
\end{equation}
where $\hat d$ is a lower bound on the minimum distance. Also notice that Algorithm~\ref{alg:new simulation approach} can be implemented in conjunction with any efficient decoder, which usually results in an upper bound on the conditional error probability. As a final remark, we need point out that the proposed simulation approach can also apply to the BER evaluation, where $f(r)$ in Step~1) of Algorithm~\ref{alg:new simulation approachforallr} is replaced by the conditional bit-error probability $f_b(r)$ and $A_{d_{\min}}$ in Step 2)
and 3) of Algorithm~\ref{alg:new simulation approachforallr} is replaced by $\sum\limits_{i = 1}^{d_{\min}}  \frac{i}{k} A_{i,d_{\min}}$. The latter
replacement can be justified by an argument similar to the proof
of Lemma~\ref{LemmaLittleSphere}.

\section{Numerical Results}

Three numerical examples are provided in this section.


\begin{table*}
\caption{A partial list of simulated conditional error probabilities for the $[7,4,3]$ Hamming code. }\label{HammingList}
  \centering\begin{tabular}{|c||c|c|c|c|c|c|c|c|c|c|}
              \hline
             $r$   & 3.50 & 3.45 & 3.40 & 3.35 & 3.30 & 3.25 &3.20 & 3.15 &  3.10\\
             \hline
             $f(r)$ & 0.493 &0.480& 0.471 & 0.458& 0.447 &  0.435 &0.423 & 0.414 &0.394 \\
              \hline
              \hline
             $r$   & 3.05 & 3.00 &2.95  & 2.90 & 2.85& 2.80 & 2.75 &2.70 & 2.65 \\
             \hline
            $f(r)$ & 0.382 &0.369& 0.356 & 0.342& 0.323 &  0.303 &0.294 & 0.277 &0.257 \\
              \hline
             \hline
             $r$   & 2.60 & 2.55 & 2.50 & 2.45 & 2.40 & 2.35 &2.30 & 2.25 &  2.20\\
             \hline
            $f(r)$ & 0.238 &0.221& 0.200 & 0.183&  0.161 &  0.140 &0.118 & 0.0975 &0.0759 \\
              \hline
               \hline
             $r$   & 2.15 & 2.10 &2.05  & 2.00 & 1.95& 1.90 & 1.85 &1.80 &  \\
             \hline
             $f(r)$ & 0.0590 &0.0419& 0.0287 & 0.0190& 0.0113 & 0.00566 &0.00217 & 4.58$\times 10^{-4}$ & \\
              \hline
            \end{tabular}
\end{table*}

\begin{figure}
\centering
\includegraphics[width=8.5cm]{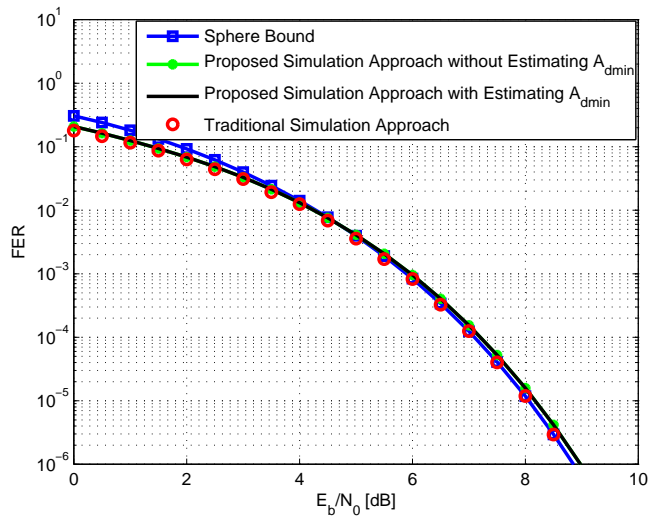}
\caption{Comparisons between different simulation approaches on FER under ML decoding: $[7,4,3]$ Hamming code.} \label{HAMMINGbound}
\end{figure}

{\bf Example~I:~}For clarity, we take the $[7,4,3]$ Hamming code as a toy example to illustrate the new simulation approach. Fig.~\ref{HAMMINGbound} shows the comparisons between the proposed simulation approach (Algorithm~\ref{alg:new simulation approachforallr}) and the traditional simulation approach (Algorithm~\ref{alg:traditional simulation approach}) on FER under ML decoding. Also shown is the original sphere bound. In the proposed simulation approach, we set $\delta_2=3.5$ and $\Delta=0.05$. For illustration, we present in Table~\ref{HammingList} a partial list of simulated conditional frame error probabilities $f(r)$. At $\delta_1 = 1.8$,  $f(\delta_1) = 4.58 \times 10^{-4}$. Since the minimum distance is $d_{\min} = 3$ and the conditional pair-wise error probability $p_2(\delta_1,d_{\min}) =6.54 \times 10^{-5}$  from (\ref{SB_f2}), we have an estimate $A_{d_{\min}} \approx 7.003$ in Step~2) of Algorithm~\ref{alg:new simulation approachforallr}. As we know, the WEF of the $[7,4,3]$ Hamming code $A(Z) = 1+7Z^3+7Z^4+Z^7$, validating the
effectiveness of the proposed estimation. It can be seen from Fig.~\ref{HAMMINGbound} that the proposed simulation approach matches well with the traditional
simulation approach.


{\bf Remark:~}It is not necessary to estimate $A_{d_{\min}}$ in this example since $\delta_1 = 1.8$ is already close to the boundary $r_0 = \sqrt{3}$. We can implement Algorithm~\ref{alg:new simulation approach} for one more sphere with radius $r_1 = 1.75$, resulting in $f(r_1) = 9.38 \times 10^{-6}$.  Given all these simulated conditional error probabilities $f(r_{\ell})$, we can have a performance curve according to (\ref{NEWBound}) in Theorem~\ref{Theorem_NEW}. This curve, as expected, coincides with the curve obtained by involving the estimated $A_{d_{\min}}$, as shown in Fig.~\ref{HAMMINGbound}.

\begin{table*}
\caption{A partial list of simulated conditional error probabilities for the $[2004,1000,5]$ terminated convolutional code. }\label{ConvList}
  \centering\begin{tabular}{|c||c|c|c|c|c|c|c|c|c|}
           \hline
             $r$    & 35.0   & 34.5  & 34.0  & 33.5 & 33.0   & 32.5  & 32.0  & 31.5 \\
             \hline
             $f(r)$  & 0.989  & 0.982 & 0.969  & 0.954  & 0.926 &  0.877 & 0.840 &  0.750\\
             \hline
              \hline
             $r$  & 31.0 & 30.5 & 30.0   & 29.5 & 29.0  & 28.5  & 28.0 & 27.5 \\
             \hline
             $f(r)$ & 0.695   &   0.632 & 0.550  & 0.459 & 0.365  &0.313 & 0.260  &0.200\\
              \hline
             \hline
             $r$     & 27.0&26.5 & 26.0  & 25.5  & 25.0  & 24.5 & 24.0  & 23.5 \\
             \hline
             $f(r)$    & 0.161& 0.115 & 0.0846   & 0.0622  &0.0429  & 0.0305 &0.0210    & 0.0147\\
                 \hline
                 \hline
             $r$   & 23.0  & 22.5&22.0 & 21.5  & 21.0     & 20.5 & 20.0 & 19.5   \\
             \hline
             $f(r)$ & 0.00847& 0.00521& 0.00311& 0.00192& 9.95$\times 10^{-4}$  & 5.60$\times 10^{-4}$   & 3.10$\times 10^{-4}$ & 1.40$\times 10^{-4}$  \\
              \hline
              \hline
             $r$  & 19.0  & 18.5& 18.0 &  17.5   &  17.0 &  16.5& & \\
             \hline
             $f(r)$   & 7.75$\times 10^{-5}$   & 2.94$\times 10^{-5}$&  1.36$\times 10^{-5}$ &  4.30$\times 10^{-6}$ &  1.63$\times 10^{-6}$ &  4.90$\times 10^{-7}$ && \\
              \hline
            \end{tabular}
\end{table*}

\begin{figure}
\centering
\includegraphics[width=8cm]{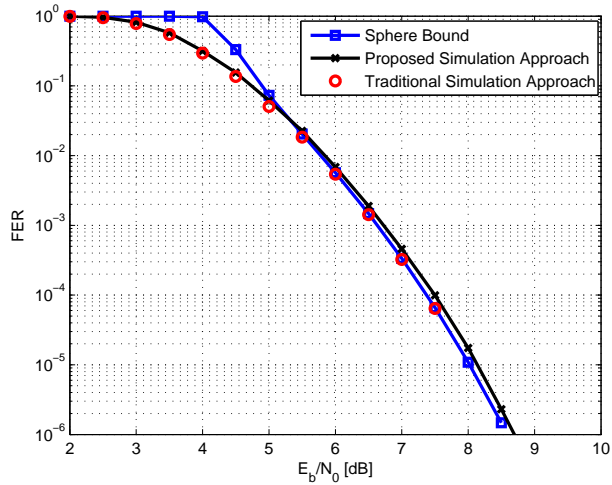}
\caption{Comparisons between different simulation approaches on FER under ML decoding: $[2004,1000,5]$ terminated convolutional code.} \label{Convbound}
\end{figure}


\begin{table*}
\caption{A list of estimated $A_{d_{\min}}$ for the $[2004,1000,5]$ terminated convolutional code.}\label{ConvAdimList}
  \centering\begin{tabular}{|c||c|c|c|c|}
              \hline
             $r$   & 17.5   &  17.0 &  16.5\\
             \hline
             $f(r)$ &  4.30$\times 10^{-6}$ &  1.63$\times 10^{-6}$ &  4.90$\times 10^{-7}$  \\
              \hline
             $p_2(r,d_{\min}) $   & 4.47 $\times 10^{-9}$ & 1.52 $\times 10^{-9}$ & 5.01 $\times 10^{-10}$ \\
             \hline
             $A_{d_{\min}}$ & 961.97 &1072.37& 978.04  \\
              \hline
            \end{tabular}
\end{table*}

\begin{table*}
\caption{A partial list of simulated conditional error probabilities for the $[961,721,\geq 32]$ QC-LPDC code. }\label{LDPCFERList}
  \centering\begin{tabular}{|c||c|c|c|c|c|c|c|c|c|c|c|}
              \hline
             $r$   & 21.0  & 20.5  & 20.0  &19.5 & 19.0 & 18.5    & 18.0   & 17.5  & 17.0 & 16.5 \\
             \hline
             $f(r)$ & 1.00   & 0.971 & 0.901  & 0.559 & 0.204  & 0.0446  &  0.00685  & 7.54$\times 10^{-4}$ &8.40$\times 10^{-5}$  & 9.00$\times 10^{-6}$ \\
              \hline
              $f_b(r)$ & 0.0666   & 0.0607&0.0511 & 0.0303& 0.0106 & 0.00241  & 3.85$\times 10^{-4}$   & 3.93$\times 10^{-5}$ &4.70$\times 10^{-6}$  & 6.23$\times 10^{-7}$ \\
              \hline
            \end{tabular}
\end{table*}

{\bf Example~II:~}Consider the convolutional code specified by the generator polynomials $[1+D+D^2, 1+D^2]$, which is terminated with information length $k = 1000$ and code length $n = 2004$. The minimum distance is $d_{\min} = 5$. Fig.~\ref{Convbound} shows the comparisons between the proposed simulation approach (Algorithm~\ref{alg:new simulation approachforallr}) and the traditional simulation approach (Algorithm~\ref{alg:traditional simulation approach}) on FER under Viterbi decoding.  Also shown is the original sphere bound.
In the proposed simulation approach, we set $\delta_2=35$ and $\Delta=0.5$. For illustration, we present in Table~\ref{ConvList} a partial list of simulated conditional frame error probabilities $f(r)$. At $\delta_1 = 16.5$,  $f(\delta_1) = 4.90 \times 10^{-7}$. Since the minimum distance is $d_{\min} = 5$ and the conditional pair-wise error probability $p_2(\delta_1,d_{\min}) =5.01 \times 10^{-10}$ from (\ref{SB_f2}), we have an estimate $A_{d_{\min}} \approx 978.04$ in Step~2) of Algorithm~\ref{alg:new simulation approachforallr}. The WEF of the terminated convolutional
code $[2004,1000,5]$ can be found in~\cite{Divsalar99} as $A(Z) = 1+1000Z^5+1997Z^6+3988Z^7 + \cdots$, which again validates
 the effectiveness of the proposed estimation. It can be seen from Fig.~\ref{Convbound} that the proposed simulation approach matches well with the traditional
simulation approach.

{\bf Remark:~}As an inherent feature of the {\em Monte Carlo} method, the estimated $A_{d_{\min}}$ varies from simulations.  For example,  we have shown in Table~\ref{ConvAdimList} three estimates of $A_{d_{\min}} \approx  \frac{f(r)}{p_2(r, d_{\min}) }$ for $r = 17.5,17,16.5$ with $f(r) < 10^{-5}$. We may use the average
$A_{d_{\min}}$~($ \approx 1004.13$) to estimate the conditional error rates on the smaller spheres, however, which makes not much difference for the performance curve.

\begin{figure}
\centering
\includegraphics[width=8cm]{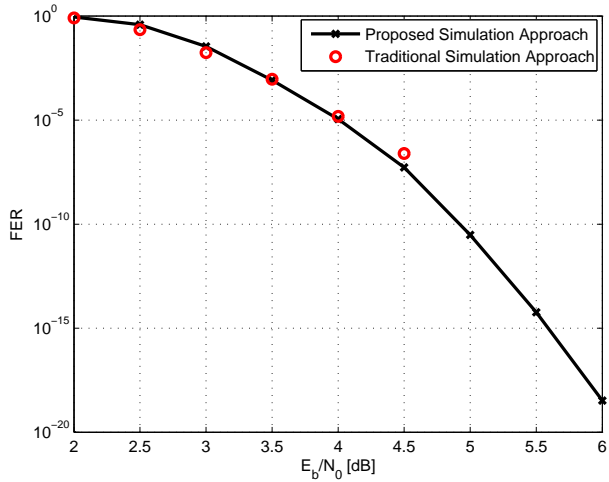}
\caption{Comparisons between different simulation approaches on FER under SPA decoding: $[961,721,d_{\min}]$ QC-LPDC code with $d_{\min} \geq 32$.} \label{LDPCFERbound}
\end{figure}

\begin{figure}
\centering
\includegraphics[width=8cm]{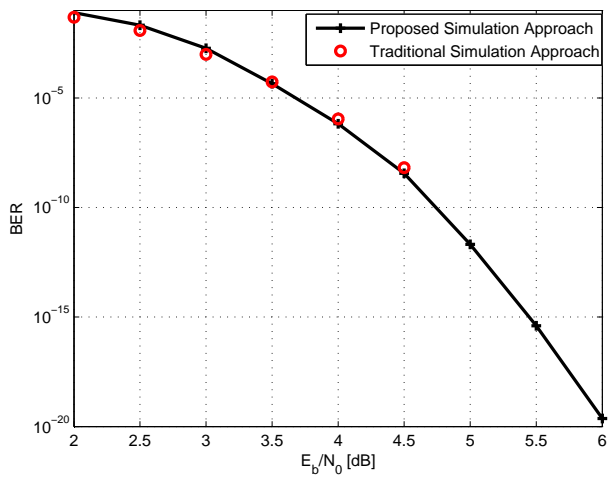}
\caption{Comparisons between different simulation approaches on BER under SPA decoding: $[961,721,d_{\min}]$ QC-LPDC code with $d_{\min} \geq 32$.} \label{LDPCBERbound}
\end{figure}

{\bf Example~III:~}Consider the $[961,721,d_{\min}]$ QC-LPDC code consturcted in~\cite{Lan07,Huang12}, where $d_{\min} \geq \hat d = 32$. Fig.~\ref{LDPCFERbound} and~\ref{LDPCBERbound} show the comparisons between the proposed simulation approach (Algorithm~\ref{alg:new simulation approachforallr}) and the traditional simulation approach (Algorithm~\ref{alg:traditional simulation approach}) on FER and BER under sum-product algorithm (SPA) decoding, respectively.  In the proposed simulation approach, we set $\delta_2=21$ and $\Delta=0.5$. Notice that the SPA is initialized by treating $r^2 / n$ as the noise variance when evaluating by Algorithm~\ref{alg:new simulation approach} the conditional error probability on ${\partial\mathcal{R}(r)}$. For illustration, we present in Table~\ref{LDPCFERList} a partial list of simulated conditional frame-error probabilities $f(r)$ and simulated conditional bit-error probabilities $f_b(r)$. At $\delta_1 = 16.5$,  $f(\delta_1) = 9.0 \times 10^{-6}$ and  $f_b(\delta_1) = 6.23 \times 10^{-7}$. Since the lower bound of the minimum distance is $\hat{d }= 32$ from~\cite{Lan07} and the conditional pair-wise error probability $p_2(\delta_1,\hat{d}) =2.80 \times 10^{-28}$ from (\ref{SB_f2}), we have an estimate $A_{d_{\min}} \approx 3.21\times 10^{22}$ and $ \sum\limits_{i = 1}^{\hat d}  \frac{i}{k} A_{i,d_{\min}} \approx 2.23 \times 10^{21}$ in Step~2) of Algorithm~\ref{alg:new simulation approachforallr}. It can be seen from Fig.~\ref{LDPCFERbound} and~\ref{LDPCBERbound} that the proposed simulation approach matches well with the traditional simulation approach in the high error region and, most importantly, is able to estimate the performance in the extremely low error region.


%


\section{Conclusions}\label{conclusion}

First, we presented in this paper a general bounding framework
based on nested partition. Second, we re-derived the SB of Herzberg and Poltyrev based on the proposed framework and showed that this bound was equivalent to the SB of Kasami~{\em et~al.}, which was rarely cited in literature. Finally, within the proposed framework, a new simulation approach was presented. The proposed simulation approach can be applied to any
codes with efficient decoding algorithms and can be used to estimate the error
rates essentially at any SNR. Our numerical results showed that the new simulation approach
matched well with the traditional simulation approach in the high error rate region where both approaches can be executed with a reasonable computational resource.

\small
\bibliographystyle{IEEEtran}
\bibliography{IEEEabrv,tzzt}

\end{document}